# The electrostatic graph algorithm: a physics-defined method for converting a time-series into a weighted complex network

**Dimitrios Tsiotas[1,2,4], Lykourgos Magafas[3], and Panos Argyrakis[3,4]**

[1]. Department of Regional and Economic Development, Agricultural University of Athens, Greece, Amfissa, Greece
[2]. Department of Planning and Regional Development, University of Thessaly, Volos, Greece
[3]. Department of Physics, University of Thessaloniki, Thessaloniki, Greece
[4]. Laboratory of Complex Systems, Department of Physics, International Hellenic University, Kavala, Greece.
[*]*Correspondence*: tsiotas@aua.gr.

**Abstract**
This paper proposes a new method for converting a time-series into a weighted graph (complex network), which builds on the electrostatic conceptualization originating from physics. The proposed method conceptualizes a time-series as a series of stationary, electrically charged particles, on which Coulomb-like forces can be computed. This allows generating electrostatic-like graphs associated to time-series that, additionally to the existing transformations, can be also weighted and sometimes disconnected. Within this context, the paper examines the structural relevance between five different types of time-series and their associated graphs generated by the proposed algorithm and the visibility graph, which is currently the most established algorithm in the literature. The analysis compares the source time-series with the network-based node-series generated by network measures that are arranged into the node-ordering of the source time-series, in terms of linearity, chaotic behaviour, stationarity, periodicity, and cyclical structure. It is shown that the proposed electrostatic graph algorithm produces graphs that are more relevant to the structure of the source time-series by introducing a transformation that converts the time-series to graphs. This is more natural rather than algebraic, in comparison with existing physics-defined methods. The overall approach also suggests a methodological framework for evaluating the structural relevance between the source time-series and their associated graphs produced by any possible transformation.

## 1. INTRODUCTION

The multidisciplinary nature of the network paradigm (Brandes et al., 2013; Barabasi, 2013; Tsiotas, 2019) introduced new directions in the time-series research and led to the emergence of the complex network analysis of time-series, which is a newly established field showing remarkable development at a multidisciplinary level (Gao et al., 2016). Scholars have recently conceptualized (Zhang and Small, 2006; Yang and Yang; 2008; Lacasa et al., 2008) that transforming a time-series into a graph can produce insights that are not visible by current time-series approaches. In general, studying the topology of a graph instead of the structure of a time-series benefits time-series analysis, because it enlarges the space of embedding, from a first-order tensor (where a time-series is embedded) to a second-order tensor (where a graph is embedded) (Tsiotas and Charakopoulos, 2020). Within this context, Zhang and Small (2006) were the first who constructed graphs from pseudo-periodic time-series, while Yang and Yang (2008) extended their approach by setting correlation coefficient thresholds to define network connectivity. Xu et al. (2008) proposed a transformation for creating graphs from time-series based on different dynamic systems. Lacasa et al. (2008) built on the intuition of considering the time-series as a landscape interpreting that two nodes are connectable when they are visible to the extent no other intermediating node obstructs its visibility. Gao and Zin (2009) proposed methods (flow pattern complex network, dynamic complex network, and fluid structure complex network) to construct complex networks from experimental flow signals, and Donner et al. (2010) introduced a recurrence method converting graphs from time-series based on the phase-space of a dynamical system. Amongst the existing ones, the natural visibility graph (NVG) algorithm (or VGA) of Lacasa et al. (2008) prevailed in the literature obviously due to its intuitive physical conceptualization (visibility, optics), and because it suggested a concise transformation, to the extent that periodic series are converted to regular graphs, random series to random graphs, and fractal series to scale-free graphs. This method has become fruitful either in developing derivative methods, such as the horizontal visibility graph of Luque et al. (2009), and the visibility expansion algorithm of Tsiotas and Charakopoulos (2020), or in motivating further research about the examination of topological properties of the visibility graph in accordance to the structure of the source time-series (Liu et al., 2010; Jiang et al., 2016; Iacobello et al. 2017).

Despite the effectiveness and popularity that it has already gained, the visibility graph algorithm of Lacasa et al. (2008) builds on a binary connectivity criterion, which leads to the development of binary connections. This conceptualization is by definition restricted in generating unweighted visibility graphs that are deescalated from the magnitude describing the source time-series. This paper introduces a method for converting a time-series into a weighted graph (complex network), which builds on another physical conceptualization originating from electromagnetism in physics. The proposed method conceptualizes a time-series as a sequence of stationary and electrically charged particles, and generates an electrostatic graph that gives a more natural, rather than algebraic transformation of time-series into graphs, in comparison to the current existing method in the literature.

## 2. THE PROPOSED ALGORITHM

By considering any node $x_i = \boldsymbol{x}(i) \in \boldsymbol{x}$ in the time-series $\boldsymbol{x}$ as a static particle of electrical charge $x_i = q_i$, we can define an (either attraction or repulsion) electrostatic force $F_{ij}$ applied between any pair of nodes $i,j$ (Fig.1), according to the inverse-square (Coulomb's) law expressed by the relation (Serway, 2004):

$$F_{ij} = k_e \cdot \frac{q_i \cdot q_j}{\left(d_{ij}\right)^2} \qquad (1),$$

where $q_i$ and $q_j$ are the charges of nodes $i,j$, $d_{ij}$ is their intermediate (spatial) distance, and $k_e$ is the Coulomb's constant. Therefore, any time-series $\boldsymbol{x}$ can be considered as a series of stationary, electrically charged, particles (i.e. the time-series nodes), on which we can compute a square matrix with the Coulomb-like forces $F(\boldsymbol{x}) = \{F_{ij} \mid i,j = 1, \ldots, n\}$.

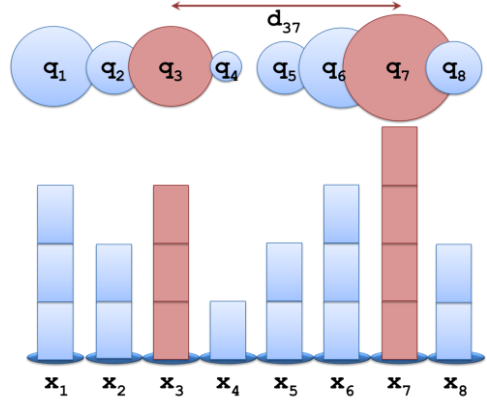

**Fig.1.** Example of the conceptualization of the electrostatic graph algorithm (ESGA). The volume of electrical charge ($q_i$, $i=1,..,8$) in each node is shown proportionally to the node size ($x_i$, $i=1,..,8$).

In the Coulomb-like forces $F(\boldsymbol{x})$ matrix, each element ($F_{ij}$) has real values expressing the (attraction or repulsion) electrostatic force $F_{ij}$ applied between any pair of nodes $i,j$. In terms of graph theory (Newman, 2010), the $F(\boldsymbol{x})$ is a square structure representing a connectivity matrix of a weighted ($w_{ij} = F_{ij}$) and undirected graph $G(V,E)$, where $V$ is the node-set and $E$ is the edge-set. However, in its current form, the Coulomb matrix $F(\boldsymbol{x})$ of the time-series represents a complete graph $K_n$ (Newman, 2010), where each node is connected with all others. In order to generate a less trivial than the complete network expressed by (the Coulomb) matrix $F(\boldsymbol{x})$, we further filter the values of $F(\boldsymbol{x})$. In particular, we define the weighted connectivity matrix $W_{ESG}$ of the associated to a time-series $\boldsymbol{x} = \{x_i \mid i=1, \ldots, n\}$ graph $G_{ESG}(V,E)$, which we will henceforth call electrostatic graph $ESG(\boldsymbol{x})$ of the time-series, as the restriction of $F(\boldsymbol{x})$ expressed by the relation:

$$W_{ESG} = \{F_{ij} \neq 0 \in F(\boldsymbol{x}) : F_{ij} \geq \left(\sum_i x_i \Big/ (n-1)\right)\} \subseteq F(\boldsymbol{x}) \tag{2},$$

where $F(\boldsymbol{x})$ is the Coulomb-like matrix of the time-series that was defined in relation (1). This connectivity criterion describes that the non-zero elements of $W_{ESG}$ (i.e. of the weighted connectivity matrix of the ESG electrostatic graph associated to $\boldsymbol{x}$) are those forces of $F(\boldsymbol{x})$ that have values higher than the adjusted mean-value $\frac{n}{n-1} \cdot \langle x \rangle$ of the time-series $\boldsymbol{x}$, where $\langle \cdot \rangle$ expresses the average operator. This quantity can be further transformed to simulate a Coulomb-like force, as follows:

$$\frac{1}{n-1} \cdot \sum_i x_i = \frac{n}{(n-1)} \cdot \frac{\sum_i x_i}{n} = \frac{n}{n-1} \cdot \langle x \rangle = n \cdot \text{sgn}(\langle x \rangle) \cdot \frac{\sqrt{|\langle x \rangle|} \cdot \sqrt{|\langle x \rangle|}}{\left(\sqrt{n-1}\right)^2} \tag{3},$$

where $\text{sgn}(\cdot)$ is the sign (or signum) function (Yun and Petkovic 2009). Within this context, the ESG connectivity criterion expresses a $n$-times stronger electrostatic force than this applied to a pair of particles with electrical charges equal to the square-root of the absolute mean-value of the time-series ($q_i$, $q_j = \sqrt{|\langle x \rangle|}$), and with distance defined by the square-root of the longest path (diameter) in the time-series ($d_{ij}=\sqrt{n-1}$). Therefore, by considering a time-series as a series of stationary and electrically charged particles, we can produce an electrostatic graph associated to the time-series, where two time-series nodes are connected whether their Coulomb-like force is greater or equal to the adjusted mean-value $\frac{n}{n-1}\cdot\langle x\rangle$ of the time-series.

Within this context, the proposed algorithm is implemented in four steps, as it is shown in Fig.2. The first computes the Coulomb-like forces matrix $F(x)$ of the time-series, the second one applies the connectivity filter to $F(x)$, the third one manages the disconnected data of $F(x)$, and the fourth step creates the graph-layout of ESG($x$).

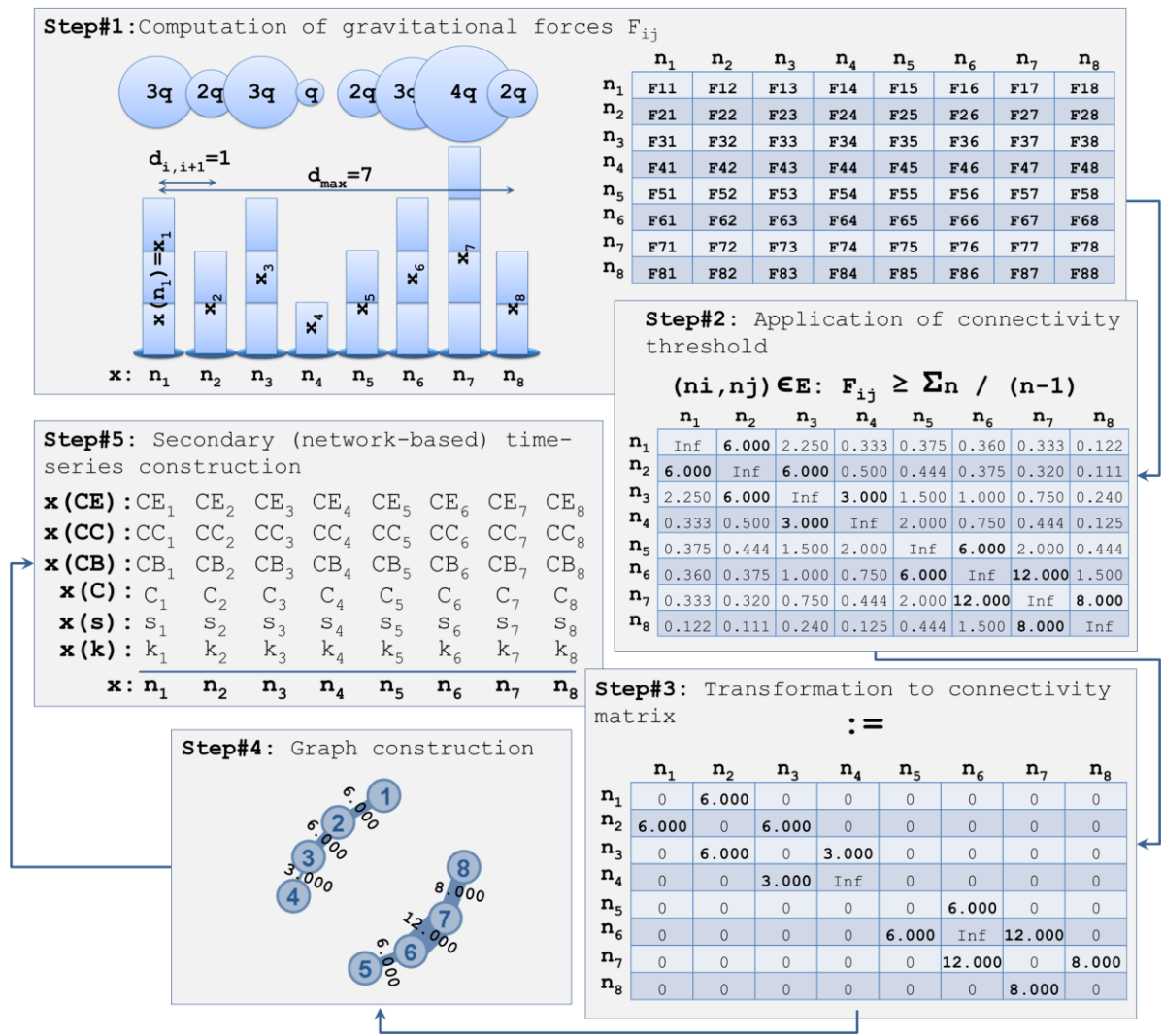

**Fig.2.** The methodological framework of the study. Steps #1 - #4 describe the ESG algorithm generating an electrostatic graph from a time-series $x = \{x_i \mid i=1, …, 8\}$. Step#5 describes the process of generating secondary time-series from the network measures of ESG($x$).

By constructing the electrostatic graph of a time-series $x$, we can further compute secondary (network-based) node-series associated to network measures the ESG($x$), because each node in the electrostatic graph $i \in V_{ESG(\boldsymbol{x})}$ corresponds to a node in the time-series $i \in V_{ESG(\boldsymbol{x})} \equiv i \in \boldsymbol{x} \mid i = 1,...,n$.

## 3. TESTING THE EFFECTIVENESS OF THE ESG ALGORITHM

The analysis builds on examining five different types of time-series, one with linear trend, another chaotic, a non-stationary, a periodic, and a cyclical. To examine the effectiveness of the proposed algorithm we firstly compare the structure of the source time-series $\boldsymbol{x}$ with its network-based node-series $\boldsymbol{x}$(ESG) generated from the ESG($\boldsymbol{x}$). This approach is expected to illustrate the level to which the topology of the associated electrostatic graph ESG($\boldsymbol{x}$) sufficiently incorporates structural information of the source time-series $\boldsymbol{x}$. Secondly, we compare the structure of the network-based node-series $\boldsymbol{x}$(ESG) generated from the ESG($\boldsymbol{x}$) with this of their homologue network-based node-series $\boldsymbol{x}$(VGA) generated from the visibility graph algorithm (VGA) of Lacasa et al. (2008). The analysis consists of a data-variability test based on Pearson's bivariate coefficient of correlation (Norusis, 2008; Walpole et al., 2012), a linear-trend test based on Linear Regression (LSLR) coefficients (Walpole et al., 2012), a chaotic-structure test based on the correlation dimension and embedding dimension diagram (Theiler, 1990; Aleksic, 1991), a stationary-structure test based on the augmented Dickey–Fuller test (ADF) for a unit root (Shumway and Stoffer, 2017), and a periodic-structure test based on autocorrelation function (Shumway and Stoffer, 2017), with reference to the behaviour of the source time-series $\boldsymbol{x}$.

To examine the effectiveness of the proposed algorithm we firstly compare the structure of the source time-series $\boldsymbol{x}$ with its network-based node-series $\boldsymbol{x}$(ESG) generated from the ESG($\boldsymbol{x}$). This approach is expected to illustrate the level to which the topology of the associated electrostatic graph ESG($\boldsymbol{x}$) sufficiently incorporates structural information of the source time-series $\boldsymbol{x}$. Secondly, we compare the structure of the network-based node-series $\boldsymbol{x}$(ESG) generated from the ESG($\boldsymbol{x}$) with this of their homologue network-based node-series $\boldsymbol{x}$(VGA) generated from the (natural) visibility graph algorithm (VGA) of Lacasa et al. (2008). The analysis detects data variability based on Pearson's bivariate coefficient of correlation (Norusis, 2008; Walpole et al., 2012), linear trend based on Linear Regression (LSLR) coefficients (Walpole et al., 2012), chaotic structure based on the correlation dimension and embedding dimension diagram (Theiler, 1990; Aleksic, 1991), stationary structure based on the augmented Dickey–Fuller test (ADF) for a unit root (Shumway and Stoffer, 2017), and periodic structure based on autocorrelation function (Shumway and Stoffer, 2017), with reference to the behaviour of the source time-series $\boldsymbol{x}$.

The proposed ESG algorithm is capable of generating weighted graphs, in contrast to VGA that exclusively generates unweighted graphs. This property is examined by the measure of network strength, which produces different network-based node-series than the measure of degree for unweighted networks. The analysis detects which network-based time series, between those generated by the proposed ESG and the VG algorithms, is closer to the behaviour of the source time-series $\boldsymbol{x}$. The results show that the ESGA appears overall more capable than the VGA in preserving the variability of the source time-series (see Supplementary Information, "Test of data variability"), it is better in preserving linearity (see SI, "Test of linearity"), and chaotic and cyclic behaviour (see SI, "Detection of chaotic structures"). In terms of stationarity, the ESGA also appears more capable than the VGA to preserve the state of stationarity of the source time-series, but the non-stationary effects of the source time-series appeared more intense (see SI, "Detection of stationarity: the Augmented Dickey–Fuller (ADF) test"). Among the available cases, the network-based node-series for the measure of strength appears closer in structure to the source time-series (Fig.3).

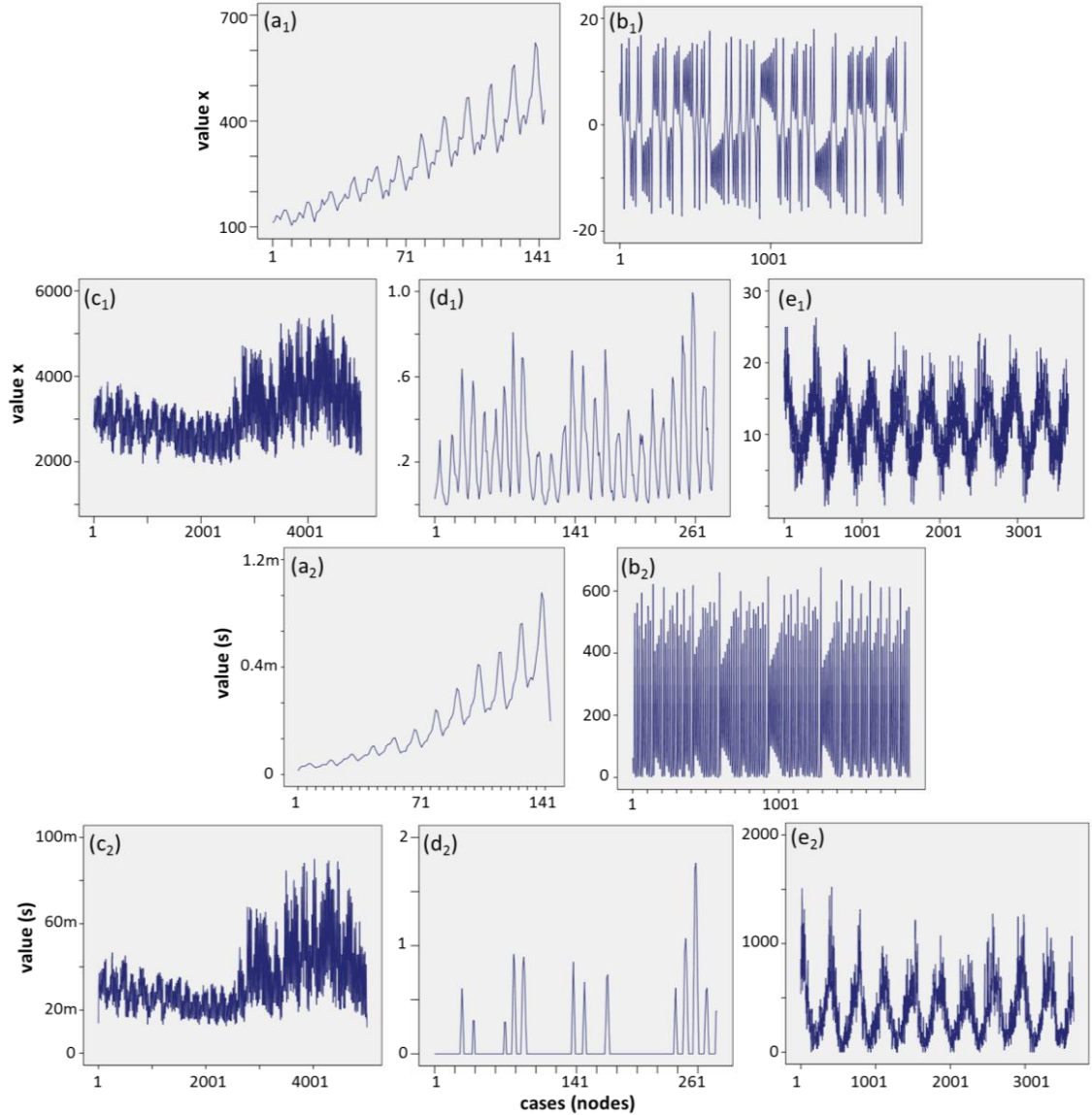

**Fig.3.** The source (reference) time-series considered in the analysis represent distinctive different patterns, where: ($a_1$) is an air-passengers time-series with linear trend ($x_a$: 144 cases, including the monthly totals of a US airline passengers for the period 1949 to 1960), ($b_1$) is the typical Lorentz chaotic time-series ($x_b$: 1900 cases, created from the Lorenz equations, on standard values sigma=10.0, r=28.0, and b=8/3), ($c_1$) is a part ($x_c$: 5000 cases) of a broader stationary time-series including estimated energy consumption, in Megawatts (MW), for the Duke Energy Ohio/Kentucky, ($d_1$) is a periodical time-series ($x_d$: 280 cases, including wolfer sunspot numbers for the period 1770 to 1771), and ($e_1$) is a cyclic time-series ($x_e$: 3650 cases, including daily minimum temperatures in Melbourne, Australia, for the period 1981-1990). The corresponding ESG network-based node-series of the measure of strength ($s$), for the available ($a_2$) air-passengers ($x_a$), ($b_2$) typical Lorentz chaotic ($x_b$), ($c_2$) DEOK ($x_c$), ($d_2$) periodical ($x_d$), and ($e_2$) cyclic ($x_e$) source time-series considered in the analysis.

Finally, the ESGA and VGA appeared equivalent in detecting periodic structure, where the ESGA outperformed in the maximum performance recorded for the measure of strength (see SI, "Detection of periodicity and cyclical structure"). Especially for the measure of strength, which is related to the immanent property of the proposed method to generate weighted graphs, the analysis shows that the network-based node-series referring to strength showed in 6 out of 9 different types of analysis carried out (5 in Table S1, one in Table S2, one in Table S3, and 2 in Table S4, see Supplementary Information) the maxima results and in 8 out 9 cases were consistent to the structure of source time-series. This highlights the added value of the proposed electrostatic graph algorithm, which attributes to the generated graphs information that is more relevant to the source time-series due to the weights included in the graph structure.

Another property of the proposed algorithm is that the ESGA can also generate disconnected graphs, whereas the VGA can only generate connected ones. Although this property is not directly related to a single measure (as the strength is directly related to the ability of producing weighted graphs), it is indirectly examined through the results observed for the ESGs corresponding to the cases of chaotic, and periodic types, which are disconnected graphs. The results show that insufficient connectivity does not restrict the ESGs to preserve structural characteristics from the source time-series since the graphs generated by the proposed method overall appeared more relevant to the initial structure of the source time-series. The authors believe that the property of insufficient connectivity indicate avenues of further research in the field of noise reduction in the time-series analysis.

## 4. CONCLUSIONS

This paper proposed a new algorithm of converting time-series to graphs by considering a time-series as a series of stationary and electrically charged particles, on which Coulomb-like forces can be computed. The proposed algorithm has an added value due to its ability to produce weighted graphs. This additional property was quantitatively examined and was found to produce graphs that are more relevant to the structure of the source time-series, implying that the proposed algorithm suggests a transformation that is more natural rather than algebraic, in comparison with the existing methods. The overall approach also suggests a methodological framework for evaluating the structural relevance between the source time-series and their associated graphs produced by any possible transformation

# SUPPLEMENTARY INFORMATION

## A. METHODOLOGY AND DATA
### ■ The conceptual framework of the algorithm

The proposed methodology builds on a joint approach combining time-series analysis and electromagnetism to convert a time-series into a complex network. In terms of probability theory and statistics (Das, 1994; Walpole et al., 2012; Shumway and Stoffer, 2017), a time-series is a vector $x$ consisting of successive nodes (real data) $x = (x_1, x_2, …, x_n)$, which are arranged into time order $t(x_1)<t(x_1)<…<t(x_n)$, from the very past to the most recent ones.

The electrostatic force applied between any pair of time-series nodes $i,j$ is defined by the expression:

$$(1) \underset{k_e \equiv 1}{\overset{q_i = x(i)}{\Leftrightarrow}} F(x(i), x(j)) = F_{ij} = \frac{x_i \cdot x_j}{(i-j)^2} \quad (S1),$$

where $(i–j)$ is the discrete distance (expressing steps of separation) between time-series nodes $i,j$ and the Coulomb's constant $k_e$ (Serway, 2004) can be considered as a scale factor, which is set in this paper to $k_e =1$. Within this context, a time-series $x = \{x_i \mid i=1, …, n\}$ can be considered as a series of stationary, electrically charged, particles (i.e. the time-series nodes) $x = \{q_i \mid q_i = x_i,\text{ for } i=1, …, n\}$, on which we can compute a square $(n \times n)$ matrix with the Coulomb-like forces $F(x) = \{F_{ij} \mid i,j =1, …, n\}$. In this matrix, each element $(F_{ij})$ has real values expressing the (attraction or repulsion) electrostatic force $F_{ij}$ applied between any pair of nodes $i,j$.

### ■ The natural visibility graph algorithm

The natural visibility algorithm (NVG) was proposed by Lacasa et al. (2008) and builds on the intuition of considering the time-series as a path of successive mountains of different height (each representing the value of the time-series at the certain time). In this time-series-based landscape, an "observer" standing on a mountain can see (either forward or backwards) as far as no other mountain obstructs its visibility (Fig.S1).

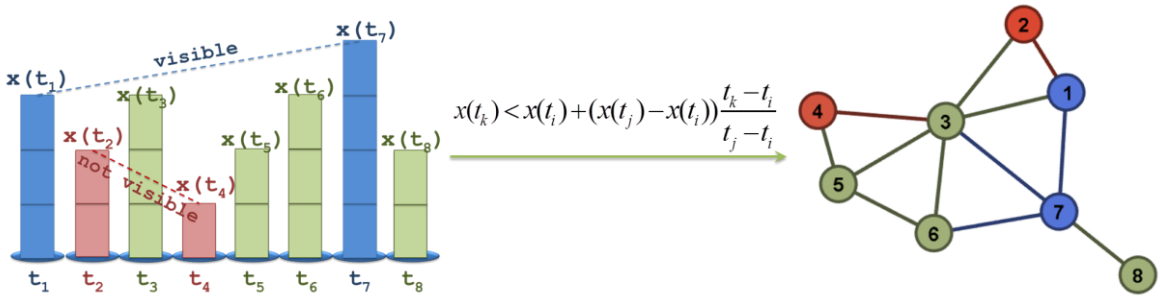

**Fig.S1. (left)** Example a pair of visible (shown in blue colour) and another of not visible (shown in red colour) time-series nodes (generally shown in green colour) defined according to the natural visibility algorithm (NVG), **(right)** the visibility graph generated from the time-series shown at the left side.

In mathematical terms, each node $(t_k, x(t_k))$ of the time-series corresponds to a graph node $n_k \equiv (t_k, x(t_k)) \in V$ (where the time-series order is preserved) and two nodes $n_i, n_j \in V$ are connected $(n_i, n_j) \in E$ when the following inequality (NVG connectivity criterion) is satisfied:

$$x(t_k) < x(t_i) + (x(t_j) - x(t_i)) \frac{t_k - t_i}{t_j - t_i} \quad (S2)$$

where $x(t_i)$ and $x(t_j)$ express the numerical values of the time-series nodes $n_i \equiv (t_i, x(t_i))$ and $n_j \equiv (t_j, x(t_j))$ and $t_i$, $t_j$ express their time-reference points.

In geometric terms, a visibility line can be drawn between two time-series' nodes $n_i$, $n_j \in V$ if no other intermediating node $n_k \equiv (t_k, x(t_k))$ obstructs their visibility, namely, no other intermediary node is so high as to intersect the visibility line created by this pair of nodes (Fig.S1). Therefore, two time-series' nodes can enjoy a connection in the associated visibility graph if they are visible through a visibility line. The visibility algorithm conceptualizes the time-series as a landscape and produces a visibility graph associated with this landscape. The associated (to the time-series) visibility graph is a complex network where complex network analysis can be further applied (Tsiotas and Charakopoulos, 2020).

■ Network measures

The network measures considered in the analysis are shown in the following Table S0.

**Table S0**
Network measures considered in the analysis.

| Measure | Description | Mathematical Expression | Reference(s) |
|---|---|---|---|
| Node Degree ($k$) | The number of edges being adjacent to a node $i$. | $k_i = k(i) = \sum_{j \in V} \delta_{ij}$, where $\delta_{ij} = \begin{cases} 1, & \text{if } e_{ij} \in E \\ 0, & \text{otherwise} \end{cases}$ | Newman (2010) |
| Node strength ($s$) | The sum of edge weights being adjacent to a given node $i$. | $s_i = s(i) = \sum_{j \in V(G)} \delta_{ij} \cdot w_{ij}$, where $w_{ij} = w(e_{ij})$ | Newman (2010) |
| Local Clustering Coefficient ($C$) | The number of a node's connected neighbors $E(i)$, divided by the number of the total triplets $k_i(k_i-1)$ shaped by the node $i$. | $C(i) = \frac{E(i)}{k_i \cdot (k_i - 1)}$ | Newman (2010) |
| Closeness Centrality ($CC$) | Total binary distance $d(i,j)$ computed on the shortest paths originating from a given node $i$ and having destination all the other nodes $j$ in the network. This measure expresses the node's reachability in terms of steps of separation. | $CC(i) = \frac{1}{n-1} \cdot \sum_{j=1, i \neq j}^{n} d_{ij} = \bar{d}_i$ | Koschutzki et al. (2005) |
| Betweenness Centrality ($CB$) | Fraction of all shortest paths $\sigma(i)$ including a given node $i$, to the number $\sigma$ of all the shortest paths in the network. | $CB(i) = \sigma(i)/\sigma$ | Koschutzki et al. (2005) |
| Eigenvector Centrality ($CB$) | Spectral measure expressing the influence of node $i$ in the network. In the formula $N(i)$ expresses the neighborhood of node $i$, $a_{ij}$ an element of the adjacency, $x_j$ the j-th | $CE(i) = \frac{1}{\lambda} \cdot \sum_{j \in N(i)} a_{ij} \cdot x_j$ | Newman (2010) |

| Measure | Description | Mathematical Expression | Reference(s) |
|---|---|---|---|
| | component of the adjacency's eigenvector with eigenvalue equal to $\lambda$. | | |

### ■ Test of data variability

At the first step of comparisons, the (Pearson's) bivariate coefficient of correlation (Norusis, 2008; Walpole et al., 2012) are computer to detect linear correlations between the variable of the source time-series $x$ and the other available variables. This approach is expected to detect amongst the network-based {$x(k)$, $x(s)$, $x(C)$, $x(CB)$, $x(CC)$, and $x(CE)$} time-series of ESG and VGA those having patterns of variability that are more relevant to the source time-series ($x$).

### ■ Test of linearity

To detect linear trend in the available time-series, we apply linear fittings to the source time-series $x$ and to its network-based node-series {$x(k)$, $x(s)$, $x(C)$, $x(CB)$, $x(CC)$, and $x(CE)$}, originating both from ESG and VGA. According this approach (Walpole et al., 2012), a linear curve $\hat{y} = b \cdot f(x) + c$ is fitted to the time-series data that bests describes the series' variability. The curve fitting algorithm estimates the parameters $b, c$ that best fit to the observed data $y$, so that to minimize the square differences $y_i - \hat{y}_i$ (Walpole et al., 2012), according to the relation:

$$\min\left\{ e = \sum_{i=1}^{n}[y_i - \hat{y}_i]^2 = \sum_{i=1}^{n}\left[ y_i - \left(\sum b_i f_i(x) + c\right)\right]^2 \right\} \quad (S3),$$

where $y_i$ are the observed values, whereas $\hat{y}_i$ are the estimated values of the response variable $y$. Estimations are implemented by using the Least-Squares Linear Regression (LSLR) method (Walpole et al., 2012), based on the assumption that the differences $e$ (relation 5) follow the normal distribution $N(0, \sigma_e^2)$. The determination of the linear fitting model is measured by the coefficient of determination $R^2$, which is defined by the expression (Norusis, 2008; Walpole et al., 2012):

$$R^2 = 1 - \left(\sum_{i=1}^{n}(y_i - \hat{y}_i)^2\right) \Big/ \left(\sum_{i=1}^{n}(y_i - \bar{y})^2\right) \quad (S4),$$

where $\bar{y}$ is the average of the observed values of the response variable and $n$ is the number of cases (time-series nodes). The coefficient of determination expresses the amount of the variability of the response variable that is expressed by the linear model and ranges within the interval [0,1], where $R^2=1$ shows perfect linear determination (Norusis, 2008; Walpole et al., 2012). Within this context, amongst the ESG and VGA network-based node-series, those being closer to the source time-series $x$ in their determination and model specialization arte considered as more relevant to $x$ in terms of linearity.

### ■ Detection of chaotic structure

The detection of the chaotic structure of time-series builds on the pattern recognition of the correlation ($v$) vs. the embedding dimension ($m$) scatter plot ($v,m$). In particular, in Chaos theory (Theiler, 1990; Aleksic, 1991), the correlation dimension ($v$) is a measure of the dimensionality of the space occupied by a set of random points, and thus is used in determining the dimension of fractal objects and is often referred to as fractal dimension. For a time-series $x = \{x_i \mid i=1, \ldots, n\}$, the correlation integral $C(\varepsilon)$ is calculated by the expression (Magafas, 2017; Hanias, 2020):

$$C(\varepsilon) = \lim_{n \to \infty} \frac{N(\varepsilon)}{n^2} \sim \varepsilon^v \qquad (S5),$$

where $N(\varepsilon)$ is the total number of pairs of points $(x_i, x_j)$ with a distance smaller than $\varepsilon$, namely $d(x_i,x_j)=d_{i,j}<\varepsilon$. As the number of points tends to infinity ($n\to\infty$), and therefore their corresponding distances tend to zero ($d_{i,j}\to 0$), the correlation integral consequently tends to the quantity $C(\varepsilon)\sim\varepsilon^v$, where $v$ is the so-called correlation dimension. Intuitively, the correlation dimension expresses the ways to which points can be close to each other along different dimensions, and this number is expected to rise faster when the space of embedding is of higher dimension. Therefore, the correlation vs. the embedding dimension diagram ($v,m$) can provide insight about the ways in which time-series points are close to each other, as the dimensionality of the space of embedding increases (Magafas, 2017; Hanias, 2020). Within this context, amongst the ESG and VGA network-based node-series , those with the ($v,m$) diagram being closer to the source time-series $x$ are considered as more relevant to $x$ in terms of chaotic structure.

■ Detection of stationarity: the Augmented Dickey–Fuller (ADF) test

The stationarity detection of the available time-series is implemented by applying an augmented Dickey–Fuller test (ADF) for a unit root (Shumway and Stoffer, 2017). The ADF algorithm examines the null hypothesis ($H_o$) that a unit-root is present in the model's time-series data, which is expressed by the relation:

$$y_t = c + \delta t + \phi \cdot y_{t-1} + \beta_1 \cdot \Delta y_{t-1} + \ldots + \beta_p \cdot \Delta y_{t-p} + \varepsilon_t \qquad (S6),$$

where $\Delta$ is the differencing operator ($\Delta y_t = y_t - y_{t-1}$), $p$ is the (user-specified) number of lagged difference terms, $c$ is a drift term, $\delta$ is a deterministic trend coefficient, $\phi$ is an autoregressive coefficient, $\beta_i$ are regression coefficients of the lag differences, and $\varepsilon_t$ is a mean zero innovation process.

According to equation (1), the unit-root hypothesis testing is expressed as follows (Shumway and Stoffer, 2017):

$$H_o : \phi = 1 \text{ vs. } H_1 : \phi < 1 \qquad (S7)$$

and the (lag adjusted) test statistic $DFt$ is defined by the expression (Shumway and Stoffer, 2017):

$$DFt = \frac{N(\hat{\phi} - 1)}{(1 - \hat{\beta}_1 - \ldots - \hat{\beta}_p)} \qquad (S8),$$

where the uppercase symbol '^' expresses an estimator. Within this context, amongst the network-based node-series of ESG and VGA, first those satisfying the null hypothesis and then those that are closer to the source time-series $x$ in terms of their $DFt$ statistic are considered as more relevant to $x$ in terms of stationarity.

■ Detection of periodicity and cyclical structure

The periodicity testing of the available time-series is based on the autocorrelation function (ACF), which is defined as:

$$\rho(s,t) = \frac{\gamma_x(s,t)}{\sqrt{\gamma_x(s,s)\gamma_x(t,t)}} \qquad (S9),$$

where ($s,t$) are time points and $\gamma_x(s,t)$ is the autocovariance function of variable $x$ (Shumway and Stoffer, 2017). In general, the ACF measures the linear predictability of the series at time $t$ by using only the value $x_s$, at time $s$, with a time-lag $dt = t - s$. The ACF lies within the interval $-1 \leq \rho(s, t) \leq 1$ where positive coefficient values imply positive linearity and negative values a negative one.

Based on the ACF, we construct a set of ACF-variables, one referring to the source time-series $x$ and the others to the network-based node-series $x(k)$, $x(s)$, $x(C)$, $x(CB)$, $x(CC)$, and $x(CE)$ originating from ESG and VGA. Each variable includes 30 elements corresponding to ACFs of lag $dt=1,2,\ldots,30$, respectively, namely:

$$ACF(x) = \{\rho(t,t+1), \rho(t,t+2), ..., \rho(t,t+30)\} \qquad (S10).$$

By constructing these *ACF*-variables, we compute the (Pearson's) bivariate coefficient of correlation (Norusis, 2008; Walpole et al., 2012) to detect linear correlations between the *ACF*($x$) variable of the source time-series $x$ and the other variables being available. Within this context, amongst the network-based node-series originating from ESG and VGA, those being more correlated to the source time-series $x$ are considered as more relevant to $x$ in terms of periodicity and cyclical (i.e. periodic with standard oscillation height) structure.

■ Data

The proposed methodology is implemented on five different time-series of typical structures, as it is shown in Fig.4. The first one (Fig.4a) was extracted from AirPassengers (2020) and is a time-series with linear trend (abbreviated: $x_a$≡AIR), including the monthly totals of a US airline passengers for the period 1949 to 1960 (144 cases). The second one (Fig.4b) was extracted from LorentzTS (2020) and is a typical Lorentz chaotic time-series (abbreviated: $x_b$≡CHAOS) created from the Lorenz equations, on standard values sigma=10.0, r=28.0, and b=8/3. This time-series has 1900 cases.

The third time-series (Fig.4c) was extracted from DEOK.hourly (2020) and is a part (the first 5000 cases) of a broader stationary time-series (of 57739 cases) including estimated energy consumption, in Megawatts (MW), for the Duke Energy Ohio/Kentucky (abbreviated: $x_c$≡DEOK). Next, the fourth one (Fig.4d) was extracted from Wolfer-sunspot-numbers (2020) and is a periodic time-series including wolfer sunspot numbers (abbreviated: $x_d$≡SUNSPOTS), for the period 1770 to 1771 (280 cases). The final time-series (Fig.4e) was extracted from Daily-minimum-temperatures-in-me (2020) and is a cyclical time-series including daily minimum temperatures in Melbourne, Australia (abbreviated: $x_e$≡TEMP), for the period 1981-1990 (3650 cases). Links of the time-series databases are available in the reference list.

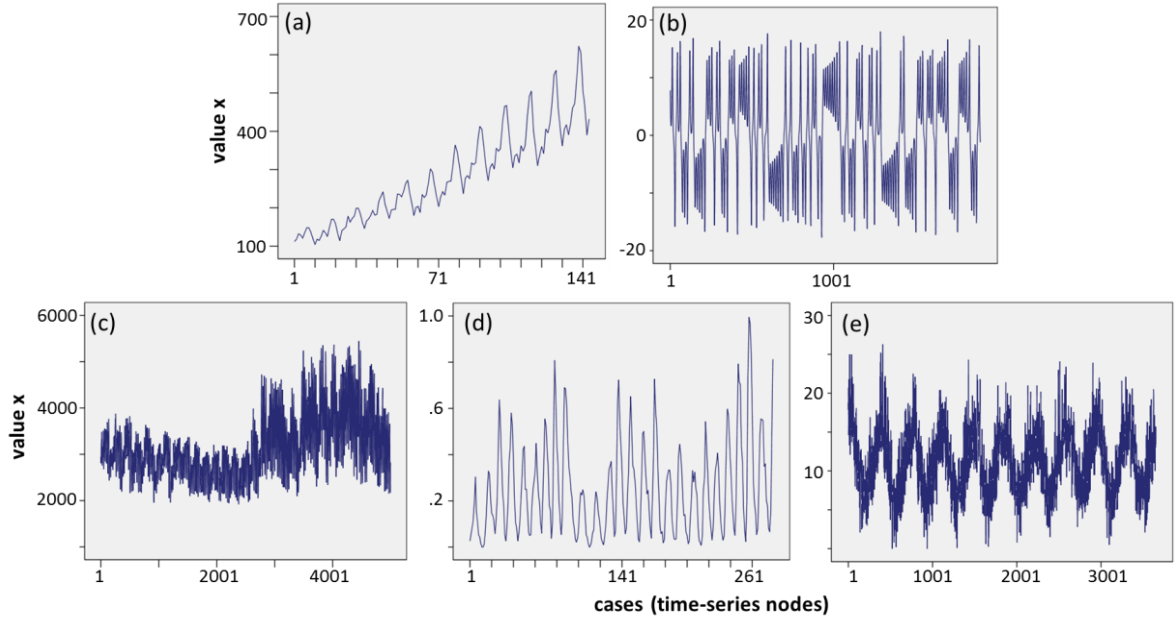

**Fig.S2.** The source (reference) time-series considered in the analysis represent distinctive different patterns. In particular, (a) is an air-passengers time-series with linear trend ($x_a$: 144 cases, including the monthly totals of a US airline passengers for the period 1949 to 1960), (b) is the typical Lorentz chaotic time-series ($x_b$: 1900 cases, created from the Lorenz equations, on standard values sigma=10.0, r=28.0, and b=8/3), (c) is a part ($x_c$: 5000 cases) of a broader stationary time-series including estimated energy consumption, in Megawatts (MW), for the Duke Energy Ohio/Kentucky, (d) is a periodical time-series ($x_d$: 280 cases, including wolfer sunspot numbers for the period 1770 to 1771), and (e) is a cyclic time-series ($x_e$: 3650 cases, including daily minimum temperatures in Melbourne, Australia, for the period 1981-1990).

## B. RESULTS AND DISCUSSION

The spy plots and graph layouts of the ESG($x$) and VGA($x$) graphs generated from the source time-series ($x$) are shown in Fig.S3-S8. The first (spy plots) are matrix-plots displaying non-zero elements of the adjacency with dots and they can thus illustrate a representation of the graph topology in the matrix-space (Tsiotas, 2019, 2020). On the other hand, the graph layout used for network visualization is the "Force-Atlas" that is available in the open-source software of Bastian et al. (2009). This layout is generated by a force-directed algorithm applying repulsion strengths between network hubs while it arranges hubs' connections into surrounding clusters. Graph models represented in this layout have therefore their hubs centered and mutually distant (where their distance between is the highest possible), whereas lower-degree nodes are placed as closely as possible to their hubs (Tsiotas, 2019).

As it can be observed in Fig.S3, the spy plot of ESG($x_a$) has a connectivity pattern configuring a tie (along the main diagonal) of increasing width (Fig.S3a), which appears indicative to the increasing trend of the source time-series ($x_a$=AIR). An aspect of such trend is also evident in the chain-like graph layout of ESG($x_a$), where a cluster of hubs appears on the right side (Fig.S3b) that resembles to the tie configuration shaped in the spy plot. However, the periodic pattern of the source time-series appears smoother in the pattern of ESG($x_a$) spy plot. On the other hand, the spy plot of the VGA($x_a$) configures a periodic pattern (Fig.S3c) where no linear trends are visible. This can be also observed in the graph layout of VGA($x_a$), which shapes an almost symmetric hub-and-spoke pattern (Fig.S3d).

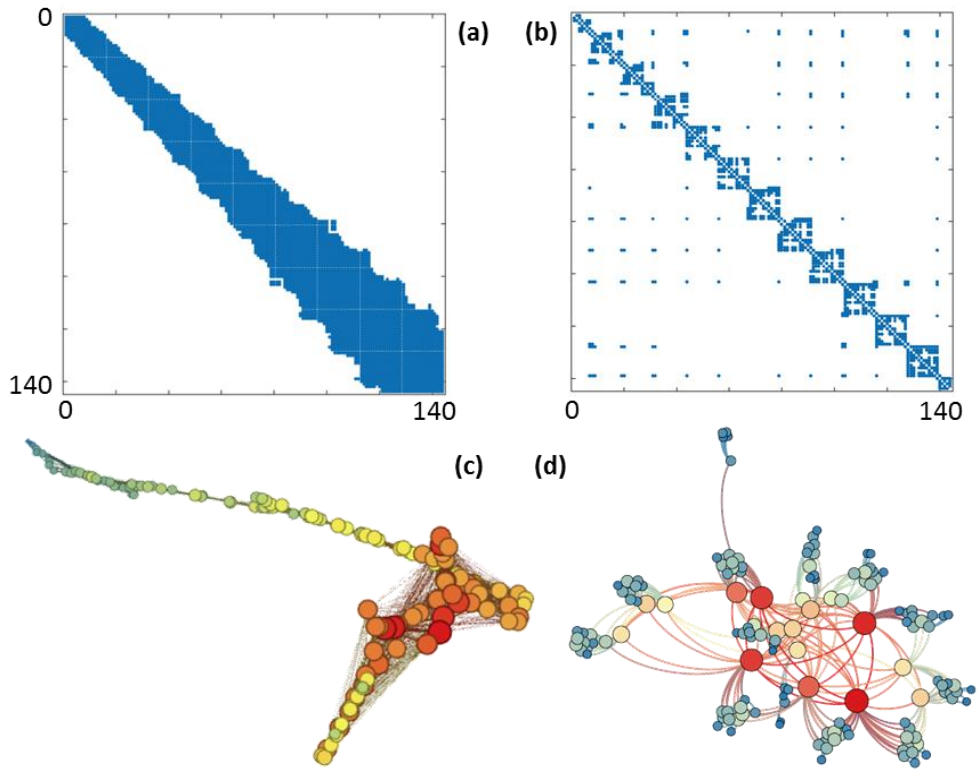

**Fig.S3.** (a) Spy plot of the ESG adjacency matrix, (b) Spy plot of the NVG adjacency matrix, (c) Force-atlas (Bastian et al. 2009) layout of the ESG, and (d) Force-atlas layout of the NVG. Both ESG and NVG are computed on the air-passengers ($x_a$) time-series (Fig.3a).

In Fig.S4, the spy plot of ESG($x_b$) configures a fractal-like tiling (Fig.S4a) illustrating a chaotic structure. Although such structure in the graph layout of ESG($x_b$) is not that clear, it is interesting to observe two major components (Fig.S4b) composing the electrostatic graph of $x_b$ (Lorentz time-series). This is a result of the positive and negative values in the structure of the source time-series ($x_b$) illustrating the ability of the electrostatic graph algorithm (ESGA) to generate disconnected graphs. Although connectivity is a desired property in complex networks (Boccaletti et al., 2006; Barabasi, 2013; Tsiotas, 2020), the ability of ESGA to generate graphs with disconnected components can be insightful for removing past or unnecessary information (noise) of the time-series and it proposes avenues for further research. On the other hand, the graph layout (Fig.S4d) of VGA($x_b$) appears more indicative to a chaotic structure than its spy plot (Fig.S4c), which is more illustrative to periodicity than to chaos.

Next, in Fig.S5, the spy plot of ESG($x_c$) configures a tie (along the main diagonal) with an almost constant width (Fig.S5a), which complies with the stationary structure of the source time-series ($x_c$=DEOK). Indications of stationarity can be also observed in the concentrated (solid-like) pattern shown in graph layout (Fig.S5b) of ESG($x_c$). On the other hand, neither the spy plot (Fig.S5c) nor the graph layout (Fig.S5d) of VGA($x_c$) are illustrative of a stationary structure, which describes the source time-series ($x_c$).

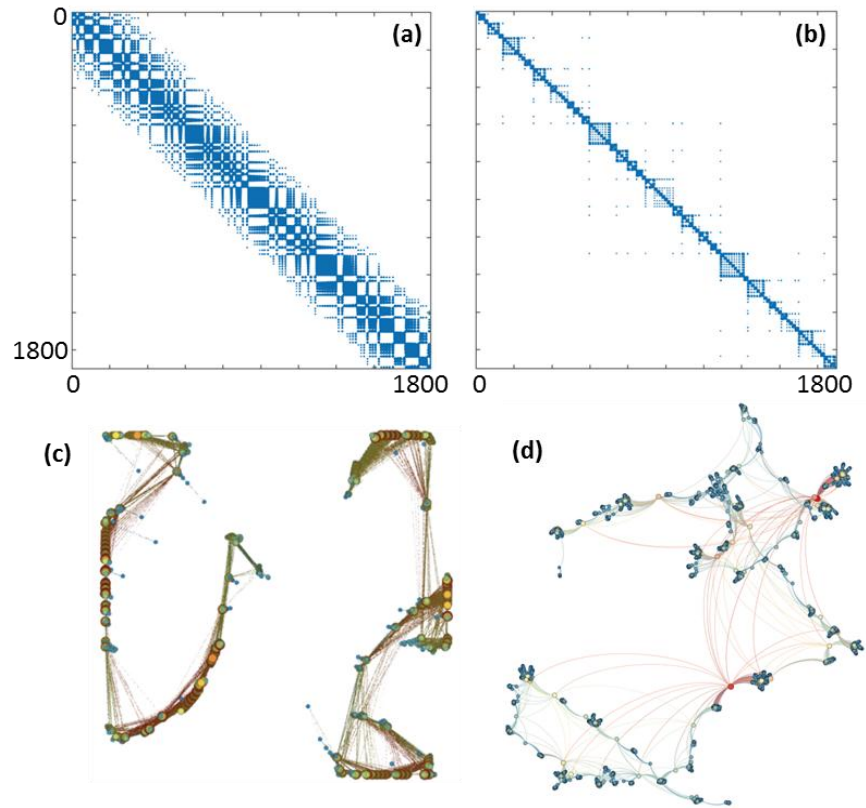

**Fig.S4.** (a) Spy plot of the ESG adjacency matrix, (b) Spy plot of the NVG adjacency matrix, (c) Force-atlas (Bastian et al. 2009) layout of the ESG, and (d) Force-atlas layout of the NVG. Both ESG and NVG are computed on the Chaos ($x_b$) time-series (Fig.3b).

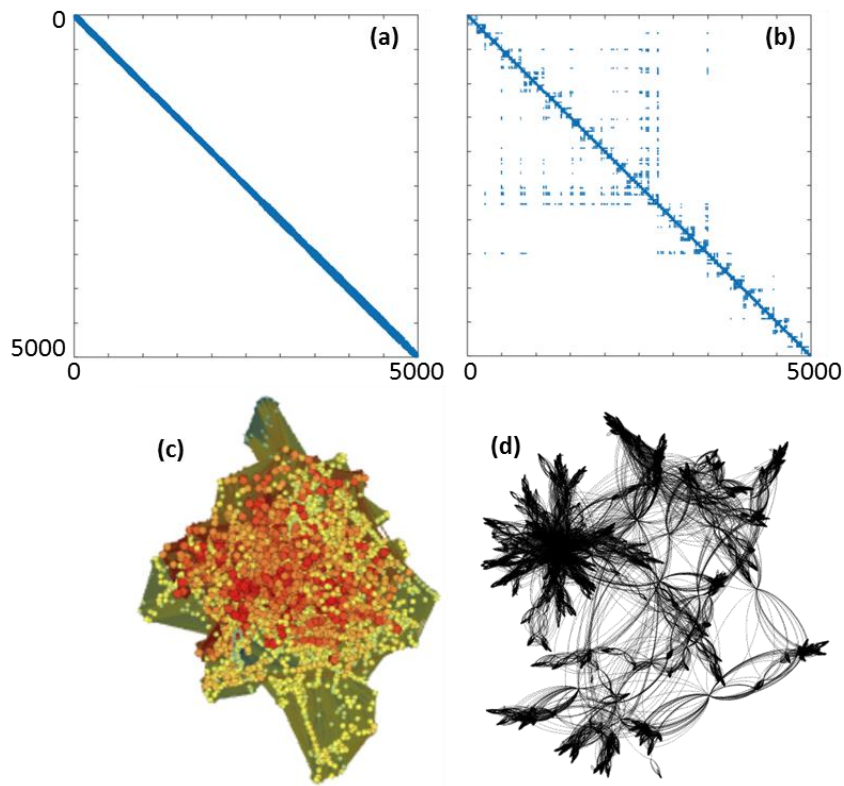

**Fig.S5.** (a) Spy plot of the ESG adjacency matrix, (b) Spy plot of the NVG adjacency matrix, (c) Force-atlas (Bastian et al. 2009) layout of the ESG, and (d) Force-atlas layout of the NVG. Both ESG and NVG are computed on the DEOK ($x_c$) time-series (Fig.3c).

In Fig.S6, the spy plot of ESG($x_d$) also configures a tie (along the main diagonal) with repeated knot-concentrations (Fig.S6a), which complies with the periodic structure of the source time-series ($x_d$=SUNSPOTS). Insightful indications of such periodicity can be also observed in the clustered (torus-like) pattern shown in the graph layout (Fig.S6b) of ESG($x_d$). On the other hand, the spy plot (Fig.S6c) of VGA($x_c$) has and interesting periodic pattern, which is slightly downgraded by the square-like dispersion with uneven range of connections appearing in the sparsity plot. However, the graph layout (Fig.S6d) of VGA($x_d$) does not appear illustrative of the periodic structure describing the source time-series ($x_d$).

Finally, the spy plot of ESG($x_e$) configures a tie (along the main diagonal) with repeated slightly thicker segments (Fig.S7a), which can relate to the cyclical structure describing the source time-series ($x_e$=TEMP). However, such cyclical structure is almost hidden in the pattern of chain graph components that have an odd arrangement in the graph layout (Fig.S7b) of ESG($x_d$). Periodicity can be clearer whether the layout will be further stretched to succeed a symmetric arrangement similar to this of Fig.S6b. On the other hand, the spy plot of VGA($x_c$) shapes a clearer periodic pattern (Fig.S7c), which can be observed (although with difficultly) in the graph layout (Fig.S7d). Overall, at a first glance, the proposed ESGA appears at least as capable as the VGA in generating graphs of topologies representative to their source time-series. This observation will be quantitatively tested at the following sections.

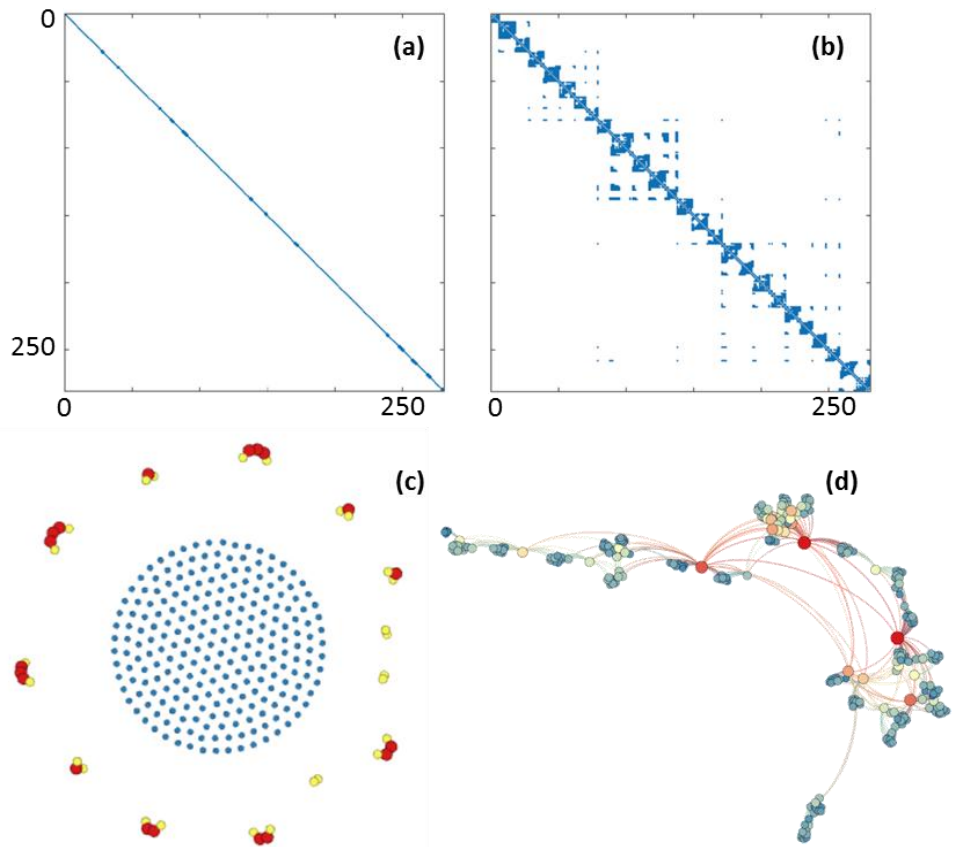

**Fig.S6.** (a) Spy plot of the ESG adjacency matrix, (b) Spy plot of the NVG adjacency matrix, (c) Force-atlas (Bastian et al. 2009) layout of the ESG, and (d) Force-atlas layout of the NVG. Both ESG and NVG are computed on the sun spots ($x_d$) time-series (Fig.3d).

Page | 9

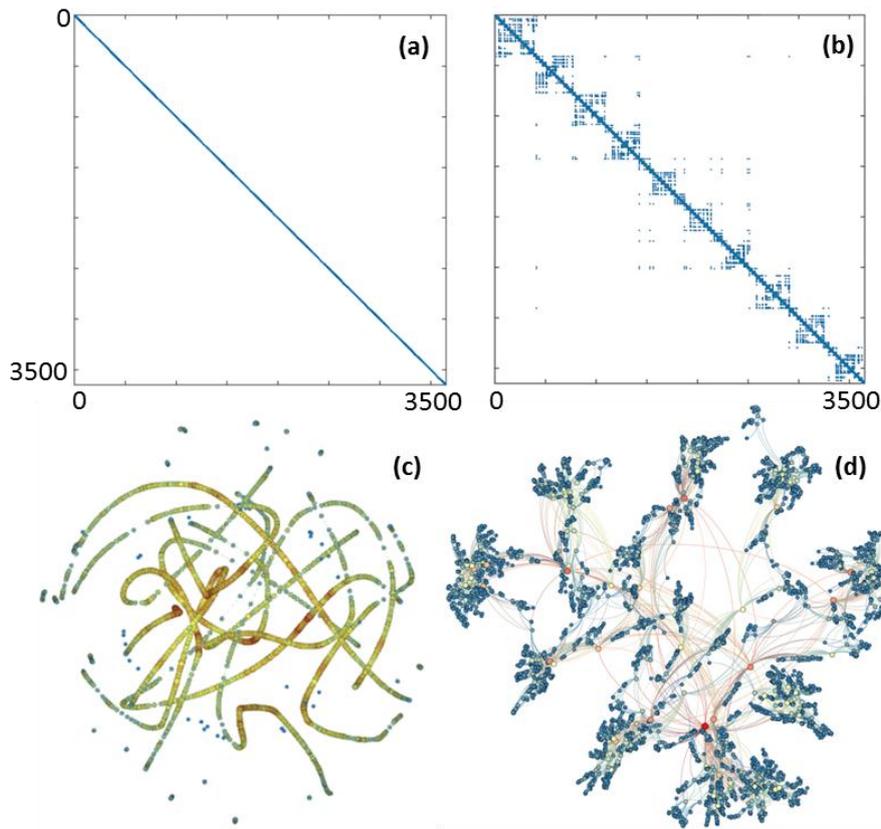

**Fig.S7.** (a) Spy plot of the ESG adjacency matrix, (b) Spy plot of the NVG adjacency matrix, (c) Force-atlas (Bastian et al. 2009) layout of the ESG, and (d) Force-atlas layout of the NVG. Both ESG and NVG are computed on the temperature ($x_e$) time-series (Fig.3e).

■ Test of data variability

To compare patterns in data variability between the source $\{x_i \mid i=a,b,\ldots,e\}$ and the network-based $\{x_i(k), x_i(s), x_i(C), x_i(CB), x_i(CC), x_i(CE) \mid i=a,b,\ldots,e\}_{ESG}$, $\{x_i(k), x_i(C), x_i(CB), x_i(CC), x_i(CE) \mid i=a,b,\ldots,e\}_{VGA}$ time-series (see Fig.S8-S12), we apply a Pearson's bivariate correlation analysis, as it is shown in Table S1. Results for the network-based node-series of strength are compared with the corresponding degree, due to the inability of VGA to produce weighted networks. As it can be observed, in the case of $x_a$ (AIR), the variability of ESGs time-series is closer to the source time-series ($x_1$) than the variability of VGAs time-series are, as being evident by the greater number (5 out of 6) in (absolutely) maximum values of correlation coefficients recorded for the ESGs. On the contrary, the variability of VGAs is closer (there are 5 out of 6 absolute maximums, where degree is counted as a double case because it equals to strength) to this of $x_b$ (CHAOS) than the variability of ESGs, except for the clustering coefficient, which in both cases has insignificant values.

**Table S1**

Results of the Pearson's bivariate correlation analysis.

| x-variable (source time-series) | Measure | y-variable (network-based node-series) | | | | | | | | | | |
|---|---|---|---|---|---|---|---|---|---|---|---|---|
| | | VGA | | | | | ESG | | | | | |
| | | $x_i(k)$ | $x_i(C)$ | $x_i(CB)$ | $x_i(CC)$ | $x_i(CE)$ | $x_i(k)$ | $x_i(s)$[c] | $x_i(C)$ | $x_i(CB)$ | $x_i(CC)$ | $x_i(CE)$ |
| $i=a$ (AIR) | r | .331** | -.158 | **.250**** | -.232** | **.254**** | **.805**** | <u>**.981****</u> | **.358**** | -.144 | **-.359**** | **.837**** |
| | sig.[a] | .000 | .059 | **.002** | .005 | **.002** | **.000** | **.000** | **.000** | .085 | **.000** | **.000** |
| | n[b] | 144 | | | | | | | | | | |
| $i=b$ (CHAOS) | r | <u>**.516****</u> | .019 | **.188**** | **-.185**** | **.201**** | -.008 | .002 | **-.022** | .020 | .103** | -.132** |
| | sig. | **.000** | .401 | **.000** | **.000** | **.000** | .712 | .919 | **.345** | .375 | .000 | .000 |
| | n | 1900 | | | | | | | | | | |
| $i=c$ (DEOK) | r | .354** | **-.570**** | .148** | -.140** | .157** | **.890**** | <u>**.989****</u> | -.549** | **.179**** | **.088**** | **.725**** |
| | sig. | .000 | **0.000** | .000 | .000 | .000 | **0.000** | **0.000** | 0.000 | **.000** | **.000** | **0.000** |
| | n | 5000 | | | | | | | | | | |
| $i=d$ (SUNSPOTS) | r | .496** | **-.666**** | .478** | -.567** | .309** | **.768**** | <u>**.773****</u> | .[b] | **.593**** | **.712**** | **.733**** |
| | sig. | .000 | **.000** | .000 | .000 | .000 | **.000** | **.000** | | **.000** | **.000** | **.000** |
| | n | 280 | | | | | | | | | | |
| $i=e$ (TEMP) | r | .437** | **-.439**** | **.211**** | **-.520**** | .164** | **.908**** | <u>**.944****</u> | .405** | .123** | -.133** | **.752**** |
| | sig. | .000 | **.000** | **.000** | **.000** | .000 | **0.000** | **0.000** | .000 | .000 | .000 | **0.000** |
| | n | 3650 | | | | | | | | | | |

a. 2-tailed significance.
b. Number of cases.
c. In pairwise consideration, the *s*(ESG) is compared with the *k*(VGA) and thus 6 pairs are configured.
**. Correlation is significant at the 0.01 level (2-tailed).
Cases shown in **bold** indicate max coefficients (in absolute terms) according to pairwise (ESG *vs*. VGA) comparisons.
<u>Underlined</u> cases indicate max coefficients (in absolute terms) within each row (for each time-series type).

In the case of $x_c$ (DEOK), the ESGs have 4 out of 6 maximum (in absolute terms) correlation coefficients, whereas the VGAs have 2 out of 6. In the case of $x_d$ (SUNSPOSTS), the respective proportions are 6 out of 6 for ESGs and 1 out of 6 for the VGAs, while in the case of $x_e$ (TEMP) these proportions are 3 out of 6 for ESGs and 3 out of 6 for the VGAs. Especially for the measure of strength (*s*) (see Fig.S13), the analysis shows that, for all types of time-series except $x_b$ (CHAOS), the network-based node-series have the highest correlation with the source time-series. Overall, this pair-wise consideration illustrates that ESGs network-based node-series have variability closer to the source time-series ($x_i$) than the VGAs, as it is illustrated by the 18 out of 30 maximum coefficients of correlation recorded for the ESGs in contrast to the 12 out of 30 maximum coefficients recorded for the VGAs.

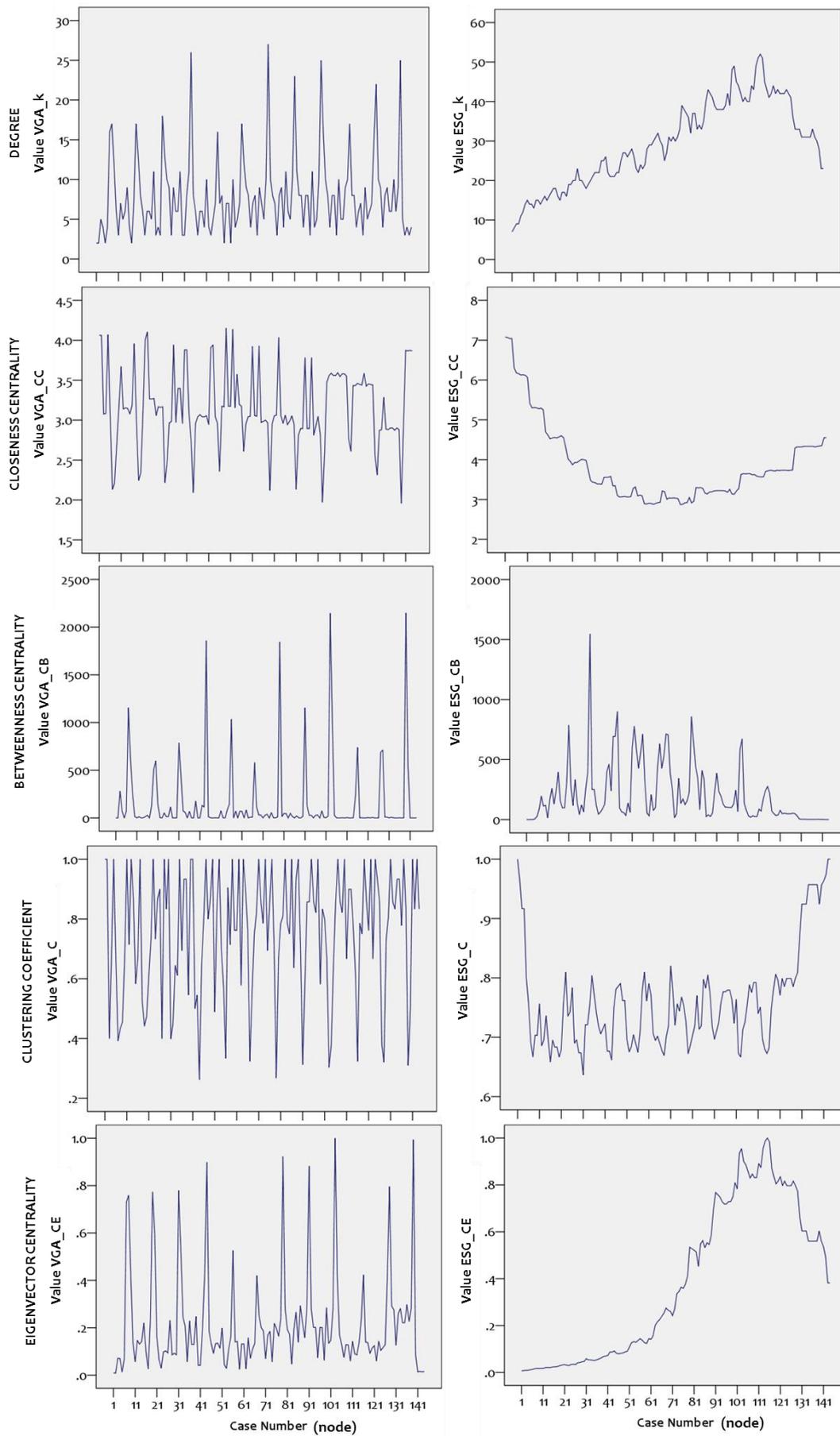

**Fig.S8.** Line plots of VGA and ESGA network-based node-series for the air ($x_a$) time-series.

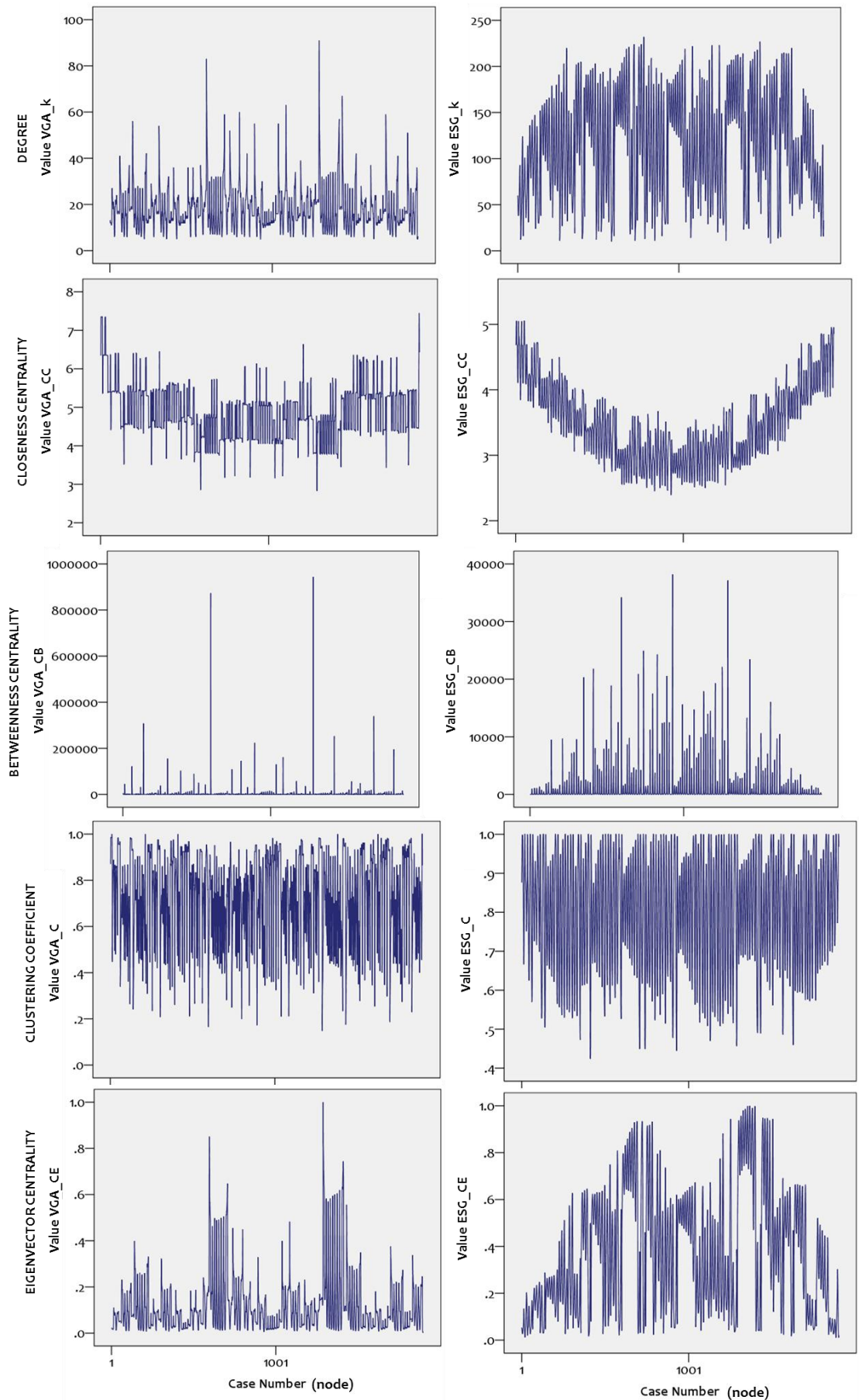

**Fig.S9.** Line plots of VGA and ESGA network-based node-series for the chaos ($x_b$) time-series.

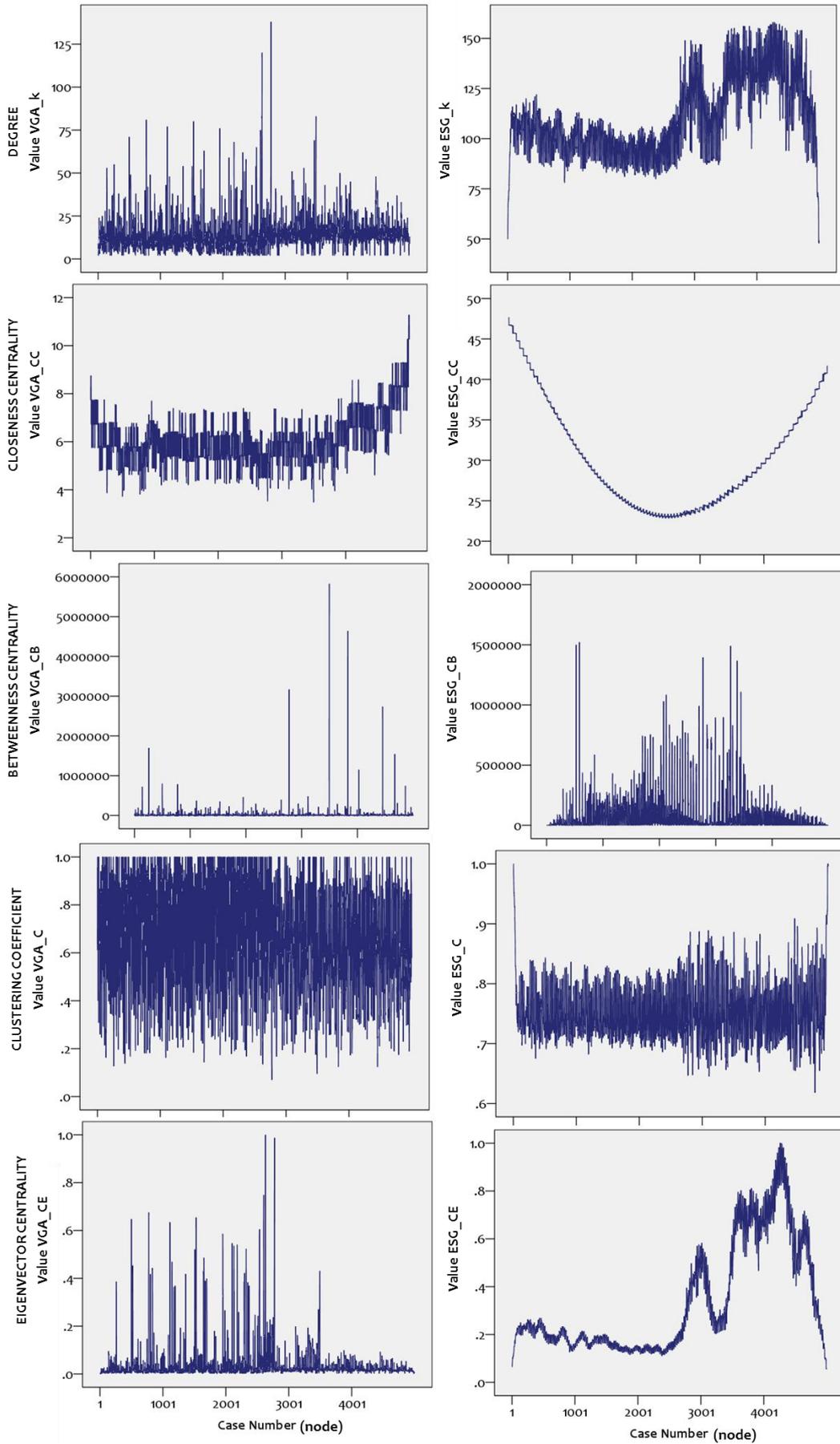

**Fig.S10.** Line plots of VGA and ESGA network-based node-series for the DEOK ($x_c$) time-series.

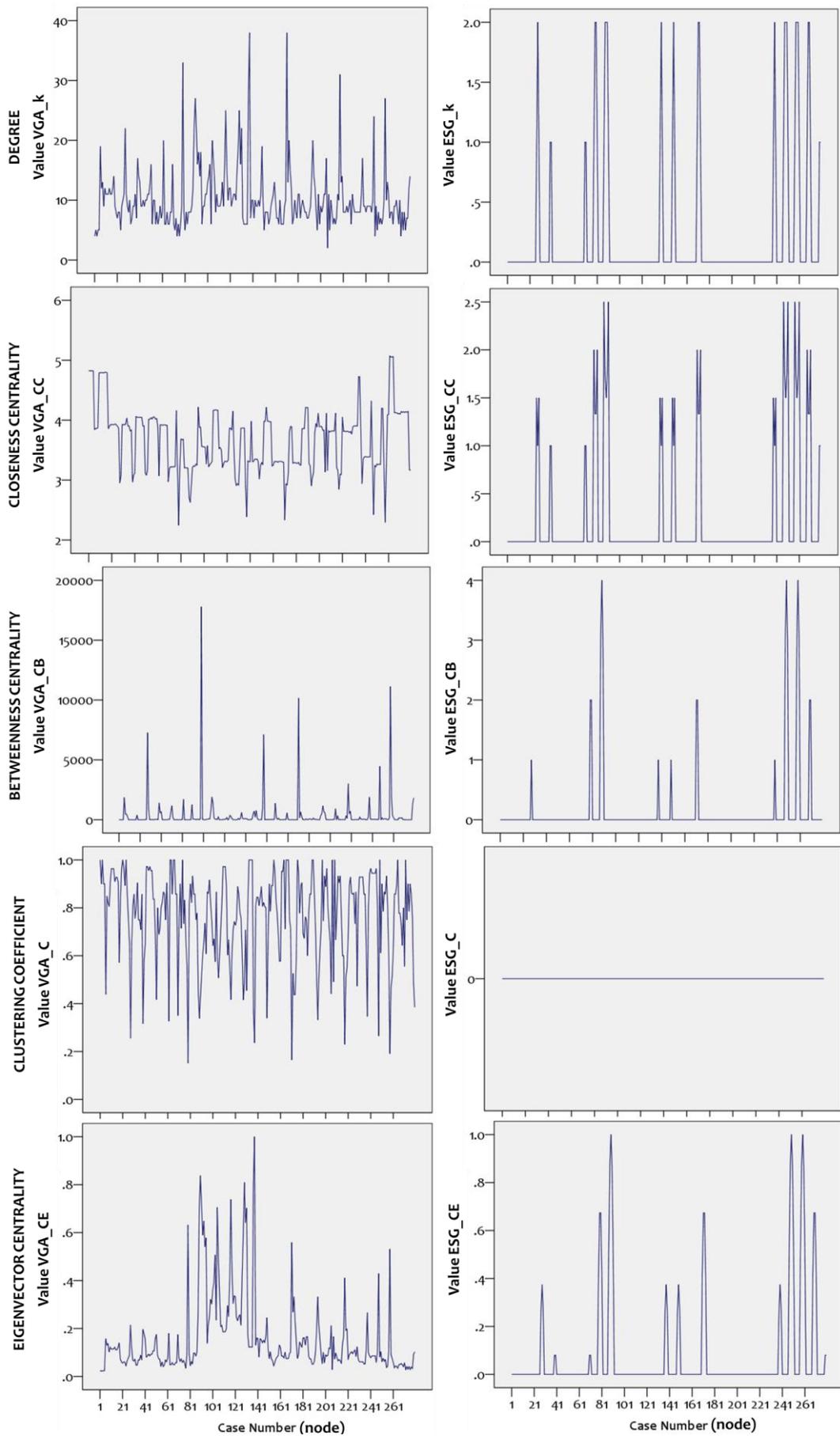

**Fig.S11.** Line plots of VGA and ESGA network-based node-series for the sunspots ($x_d$) time-series.

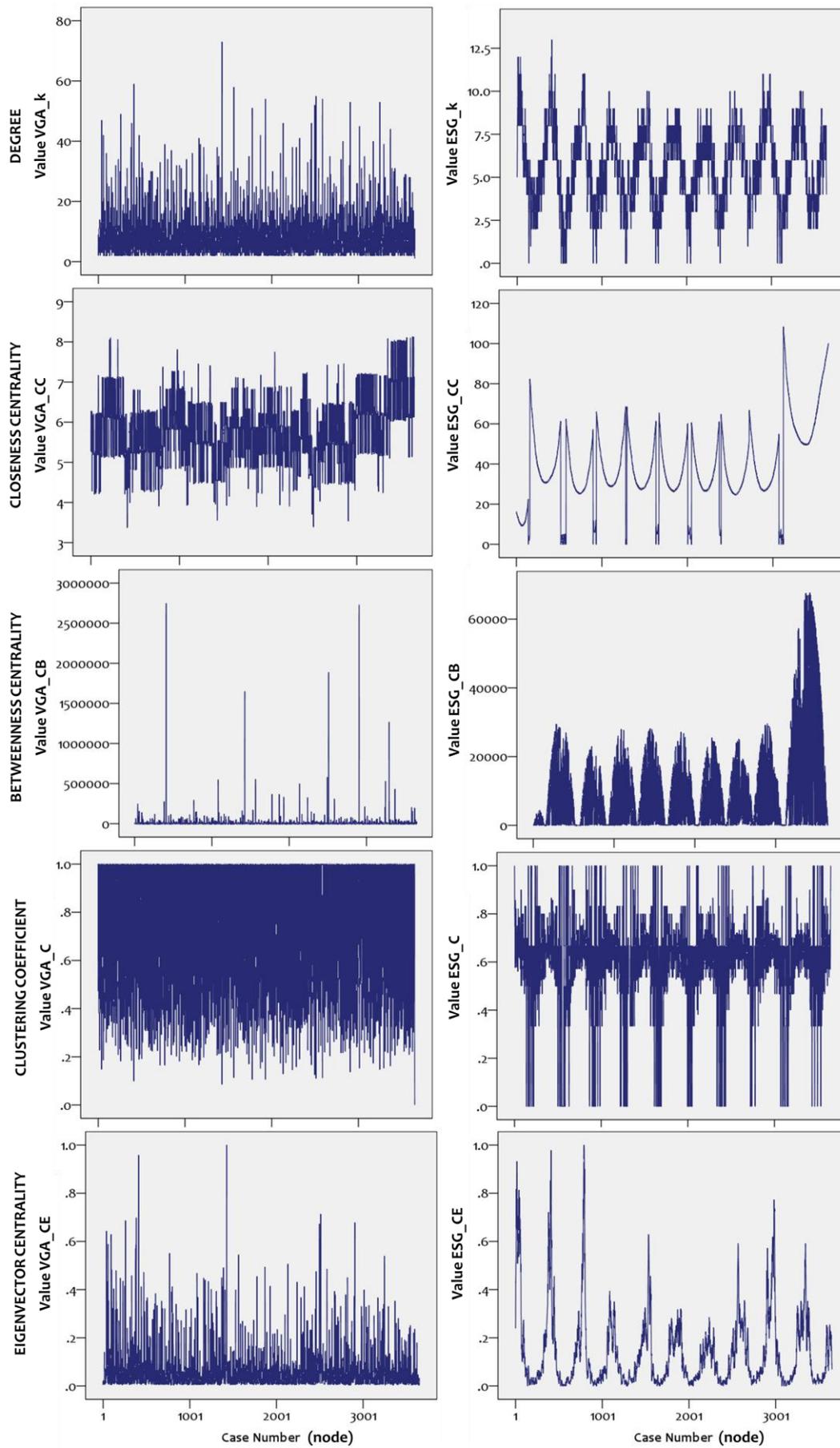

**Fig.S12.** Line plots of VGA and ESGA network-based node-series for the temp ($x_e$) time-series.

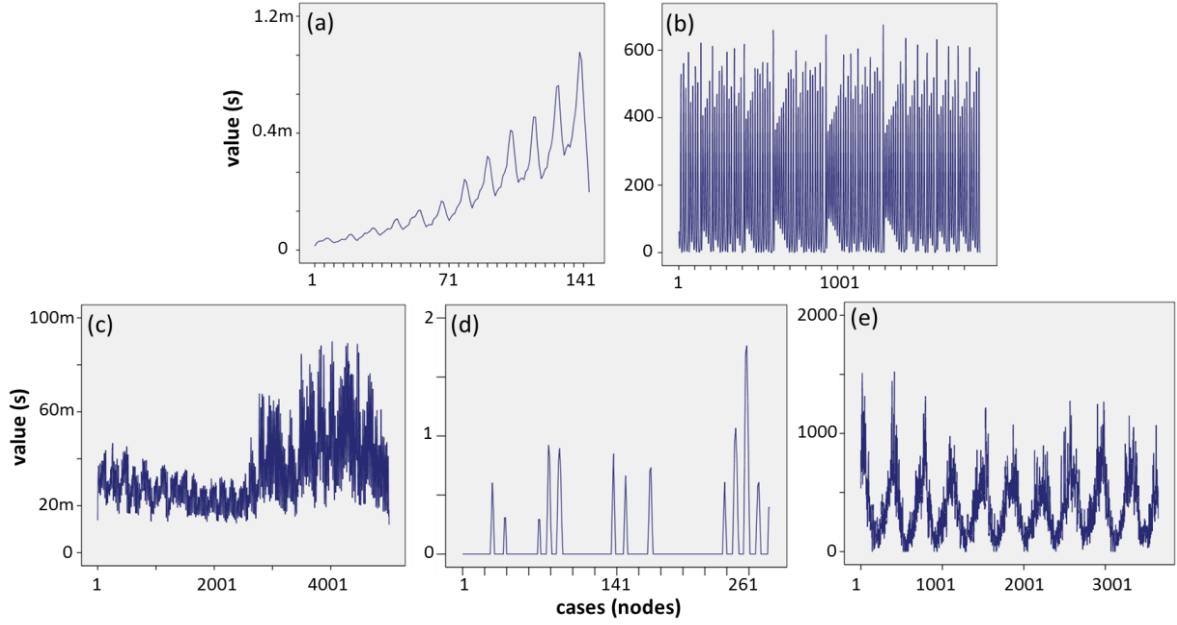

**Fig.S13.** The ESG network-based node-series of the measure of strength (*s*), for the available (a) air-passengers ($x_a$), (b) typical Lorentz chaotic ($x_b$), (c) DEOK ($x_c$), (d) periodical ($x_d$), and (e) cyclic ($x_e$) source time-series considered in the analysis.

■ Test of linearity

Linearity testing was applied to the $x_a$ (AIR) time-series, which has an obvious linear trend. The results of the analysis are shown in Table S2, where first it can be observed that the source ($x_a$: AIR) time-series is satisfactorily described by a linear model ($R^2$=0.8536). However, none of the VGAs sufficiently retain this linear structure, as it can be observed from the respective low coefficients of determination, which range between 0.0002-0.0132.

**Table S2**
Linear regression fittings for the $x_a$ (Air) time-series.

| | Time-series/Measure | Linear regression | |
|---|---|---|---|
| | | Mathematical expression | Determination |
| | $x_a$ (AIR) | $y = 2.6572x + 87.653$ | $R^2 = 0.8536$ |
| VGAs | $k(x_a)$ | $y = 0.0118x + 7.0629$ | $R^2 = 0.0096$ |
| | $C(x_a)$ | $y = 0.0006x + 0.7174$ | $R^2 = 0.0132$ |
| | $CB(x_a)$ | $y = 0.174x + 141.63$ | $R^2 = 0.0003$ |
| | $CC(x_a)$ | $y = -0.0002x + 3.1682$ | $R^2 = 0.0002$ |
| | $CE(x_a)$ | $y = 0.0001x + 0.2013$ | $R^2 = 0.0009$ |
| ESGs | $k(x_a)$ | $y = 0.2149x + 13.656$ | **$R^2 = 0.6916$** |
| | $s(x_a)$ | $y = 4813.4x – 62583$ | **$R^2 = 0.8012$** |
| | $C(x_a)$ | $y = 0.0009x + 0.6969$ | $R^2 = 0.1985$ |
| | $CB(x_a)$ | $y = -1.5134x + 315.61$ | $R^2 = 0.0663$ |
| | $CC(x_a)$ | $y = -0.0106x + 4.6484$ | $R^2 = 0.2034$ |
| | $CE$ | $y = 0.0069x - 0.1143$ | **$R^2 = 0.7579$** |

Cases shown in **bold** indicate high (> 0.65) coefficients of determination. Underlined cases highlight the max coefficient, in total.

On the contrary, the ESGs degree *k*(ESG), strength *k*(ESG), and eigenvector centrality *CE*(ESG) time-series have a considerable linear structure, as it is shown by their respective coefficients of determination $R^2=0.6916$, $R^2=0.8012$, and $R^2=0.7579$, among which strength (*s*) has the highest determination. Overall, this analysis illustrates that the ESGA appears more capable than the VGA in generating graphs that can preserve aspects of linearity of the source time-series.

■ Detection of chaotic structures

In this step of analysis the correlation *vs*. the embedding dimension diagrams (*v,m*) of the VGAs and ESGs time-series are compared for chaotic structure in reference to the chaotic time-series $x_b$ (CHAOS), which is by default constructed on the Lorenz equations. The results (Fig.S14) show that all (*v,m*) diagrams of the ESG time-series (except this of eigenvector centrality $x_b(CE/_{ESG})$) illustrate chaotic structure but mostly of different characteristics than the source chaotic time-series $x_b$. However, the (*v,m*) diagrams of strength $x_b(s/_{ESG})$ and the source time-series $x_b$ almost coincide, which implies a relevant chaotic structure between these time-series. On the other hand, degree $x_b(CE/_{VGA})$ and possibly the eigenvector centrality $x_b(CE/_{VGA})$ network-based node-series of VGA illustrate chaotic structure of high dimensionality, which are also of different characteristics than the source chaotic time-series $x_b$. Overall, the chaos analysis shows that the ESG is more capable incorporating in the network structure the chaotic structure of the source time-series. Particularly the measure of strength shows the most relevant chaotic structure which almost coincides with this of source time-series.

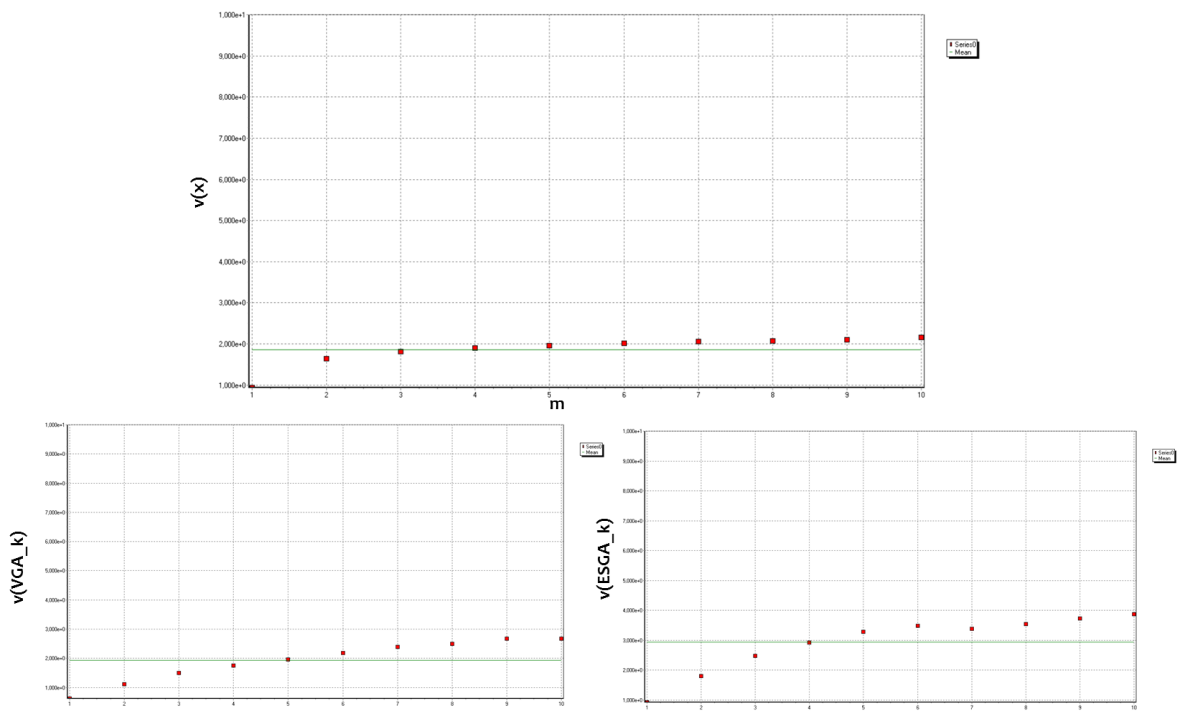

**Fig.S14 (part A: degree).** Results of the chaotic structure detection analysis applied to the typical Lorentz chaotic (source) time-series ($x_b$) and to the VGAs and ESGs network-based node-series of degree (*k*).

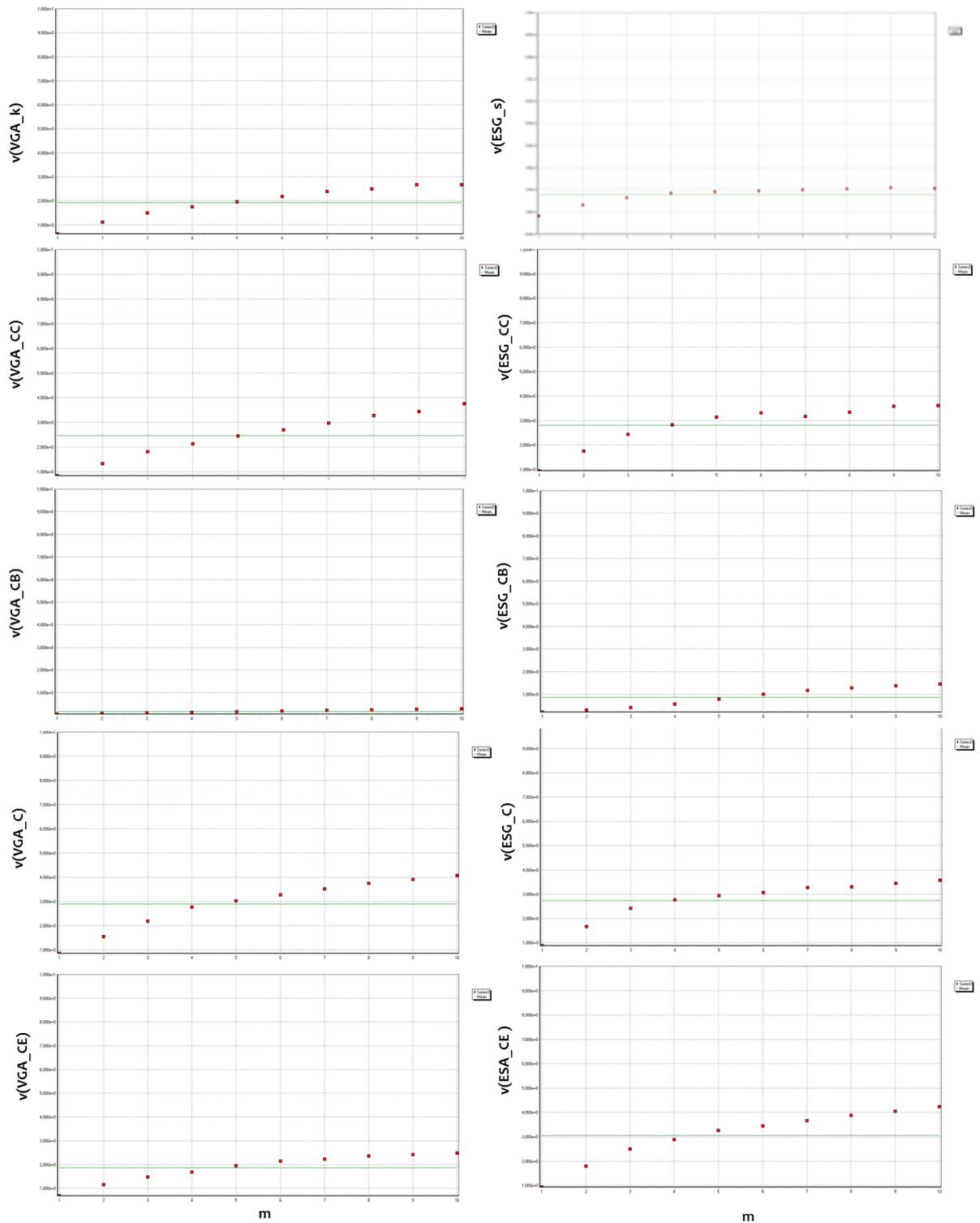

**Fig.S14 (part B).** Results of the chaotic structure detection analysis applied to the VGAs and ESGs network-based node-series of strength ($k$,$s$), closeness centrality (CC), betweenness centrality (CB), clustering coefficient (C), and eigenvector centrality (CC), which are associated to the typical Lorentz chaotic (source) time-series ($x_b$) (shown in Fig.S14.A).

■ Detection of stationarity: the Augmented Dickey–Fuller (ADF) test

Next, a stationarity test was applied to the $x_c$ (DEOK) time-series. The results of the analysis are shown in Table S3, where, first, it can be observed that is 7.03% likely for $x_c$ to have a unit-root and thus to be a non-stationary time-series. This result implies that the null-hypothesis cannot be rejected, and thus the source ($x_c$) time-series cannot be considered as stationary. The results for the VGAs are 0.1% for $k$(VGA), $C$(VGA), $CB$(VGA), and $CE$(VGA) and 1.76% for $CC$(VGA), which implies that all of these network-based time series is statistically safe to be considered as stationary.

**Table S3**

ADF test for stationarity of the $x_c$ (DEOK) time-series.

|  |  | $h^{(a)}$ | p-Value | DFt (stat) | cValue |
|---|---|---|---|---|---|
| $x_c$ (DEOK) |  | 0 | 0.0703 | -1.7876 | -1.9416 |
| VGAs | $k$ | 1 | **0.0010**[b] | -16.8555 |  |
|  | $C$ | 1 | **0.0010** | -8.8160 |  |
|  | $CB$ | 1 | **0.0010** | -63.9394 | -1.9416 |
|  | $CC$ | 1 | **0.0176** | **-2.3683** |  |
|  | $CE$ | 1 | **0.0010** | -28.4443 |  |
| ESGs | $k$ | 0 | 0.2176 | **-1.1860** |  |
|  | $s$ | 0 | 0.3876 | **-0.7177** |  |
|  | $C$ | 0 | 0.3458 | **-0.8359** |  |
|  | $CB$ | 1 | **0.0010** | -20.2674 | -1.9416 |
|  | $CC$ | 0 | 0.1739 | **-1.3179** |  |
|  | $CE$ | 0 | 0.2251 | **-1.1656** |  |

a. $h$=0 indicates failure to reject the unit-root null (indication: non-stationary).
$h$=1 indicates rejection of the unit-root null in favor of the alternative model (indication: stationary).
b. Cases shown in **bold** font are close to the source time-series scores

On the other hand, the results for the ESGs are 21.76% for $k$(ESG), 34.58% for $C$(VGA), 0.1% for $CB$(VGA), 17.39% for $CC$(VGA), and 22.51% $CE$(VGA), which implies that 4 out of 5 cases cannot be considered as stationary. An interesting observation about this result is that the (although insufficient to retain the null hypothesis) p-values of the VGAs are closer (in terms of distance) than those of the ESGs. This result implies that the non-stationary effects immanent in the source time-series are more intense in the structure of the ESGs than of the VGAs.

■ Detection of periodicity and cyclical structure

The analysis of detecting periodical and cyclical structures is based on correlations between autocorrelation variables $ACF(x)$, which are constructed by autocorrelation functions with lag 1,2,…,30, for every available network measure, according to relation (13). The results of the correlation coefficients $r(ACF(x_i), ACF(x_i|_{p,k}))$, with $i \in \{d,e\}$, $j \in \{k, s, C, CB, CC, CE\}$, and $p \in \{VGA, ESG\}$ are shown in Table S4.

**Table S4**

Correlations of ACFs[(*)] between the source and the network-based node-series (Sunspots and Temp).

| | | y-variable | | | |
|---|---|---|---|---|---|
| | | ACF($x_d$) [SUNSPOTS] | | ACF($x_e$) [TEMP] | |
| x-variable | | r | sig | r | sig |
| VGA | ACF(*k*) | **0.8397** | 0 | -0.0803 | 0.613 |
| | ACF(*C*) | 0.7953 | 0 | -0.1311 | 0.408 |
| | ACF(*CB*) | -0.082 | 0.6057 | 0.2807 | 0.0717 |
| | ACF(*CC*) | 0.4225 | 0.0053 | <u>**1**</u> | 0 |
| | ACF(*CE*) | **0.823** | 0 | 0.3306 | 0.0325 |
| ESGA | ACF(*k*) | 0.6513 | 0 | **0.9999** | 0 |
| | ACF(*s*) | <u>**0.9379**</u> | 0 | 0.9196 | 0 |
| | ACF(*C*) | NaN[**] | NaN | **0.9999** | 0 |
| | ACF(*CB*) | 0.2398 | 0.1261 | **0.9998** | 0 |
| | ACF(*CC*) | **0.7105** | 0 | 0.9999 | 0 |
| | ACF(*CE*) | 0.5601 | 0.0001 | **0.9999** | 0 |

*. *ACF*s are computed on lags from 1:30 (relation 13).
**. Not a number (unable to compute due to zero entries)
Cases shown in **bold** indicate max coefficients (in absolute terms) according to pairwise (ESG *vs*. VGA) comparisons (only statistically significant cases are shown).
<u>Underlined</u> cases indicate max coefficients (in absolute terms) within each column (for each time-series type).

For the case of $x_d$ (SUNSPOTS), 4 out of 6 (degree is counted as a double case) network-based VGA time-series (*k*, *C*, *CE*) and 3 out of 6 network-based ESG time-series (*k*, *s*, *CC*) are significantly correlated with the source time series $x_d$. Amongst these significant results, the VGAs have 2 maxima in pairwise comparisons, whereas ESGs have another 2 maxima. Moreover, the case of strength has the highest correlation coefficient amongst all available network-based node-series for the SUNSPOTS ($x_d$) typology, illustrating a better performance of the ESG algorithm to preserve periodicity due to its capability of generating weighted electrostatic networks. For the case of $x_e$ (TEMP), 1 case appears significant (closeness centrality) for the VGAs and all cases are significant for the ESGs. In terms of pairwise comparisons, the VGAs have 1 (out of 6) maximum, whereas the ESGs have 5 out of 6 maxima. However, although high, the strength does not succeed to enjoy the total maximum for the TEMP ($x_e$) case. Overall, this analysis shows that the ESGs appear more capable than the VGAs in preserving periodic and cyclical characteristics of the source time-series.

### ■ The ESGA Matlab (m-file) Code

```matlab
function [ ESGA ESGA_n fc] = esga_und( x )
%ELECTROSTATIC GRAPH ALGORITHM (ESGA) this function creates an undirected
graph associated to a time-series by using
%the electrostatic graph algorithm
%   INPUTS
%       x: a time-series vector
%
%   OUTPUTS
%       ESGA:   the associated electrostatic graph
%       ESGA_n: the complete associated electrostatic graph (i.e. prior
%               applying the electrostatic threshold)
%       fc:     the electrostatic threshold (charge), where higher than
%               (>=) fc connections are kept in the ESGA.
%
%   Developed by Dimitrios K. Tsiotas, Ph.D., 27 June 2020

tic
n=length(x);
ESGA=zeros(n);
fc=sum(x)/(n-1);
for i=1:n
    for j=1:n
    ESGA(i,j)=(x(i)*x(j))/(i-j)^2;
    end
end
ESGA_n=ESGA;
ESGA=ESGA.*(ESGA>=fc);
toc
end
```